\documentstyle[12pt]{article}
\title{\Large\bf On Fifth Order KdV-Type Equation}

\author{R.P.Malik
\thanks{E-mail address: MALIK@THSUN1.JINR.DUBNA.SU}\\
      \small \it Bogoliubov Laboratory of Theoretical Physics,\\
 \small \it JINR, 141980 Dubna, Moscow Region, RUSSIA }

\begin{document}
\hoffset=-1truecm
\voffset=-2truecm
\baselineskip=16pt
\date{}
\maketitle
\begin{abstract}
The dynamics of the highly nonlinear fifth order $KdV$-type equation is
discussed in the framework of the Lagrangian and Hamiltonian
formalisms. The symmetries of the
Lagrangian produce three commuting conserved quantities that are found to
be recursively related to one-another for a certain specific value of the
power of nonlinearity. The above cited recursion relations are obeyed
with a second Poisson bracket which  sheds light on the
integrability properties of the above nonlinear equation.
It is shown that a Miura-type transformation can be made to obtain the
fifth order $mKdV$-type equation
from the fifth order $KdV$-type equation. The spatial dependence of the
fields involved is, however, not physically interesting
from the point of view of the solitonic solutions. As a consequence,
it seems that the fifth order $KdV$- and $mKdV$-type equations
are completely
independent nonlinear evolution equations in their own right.\\

\vspace{0.3cm}

PACS number(s): 52.35.Sb, 52.35.Mw, 63.20.Ry
\end{abstract}

\pagenumbering{arabic}
\newpage
\noindent
{\Large \bf 1 Introduction} \\

\vspace{0.1cm}

\noindent
In the classical soliton theory, the ideas of integrability, elastic
scattering of solitons, stability and preservation of their shape, etc.,
are fairly well established to be very intimately connected.
In recent years, the stability
of the soliton solutions to the highly nonlinear fifth order $KdV$-type
equations has been studied extensively [1--5]. In particular, the effect
of the dispersion due to higher order derivative terms
on solitons has been studied
for the fifth order $KdV$- and $mKdV$-type equations which are useful in the
context of fluid mechanics and plasma physics [2-4]. It has been shown that
the nonzero dispersion effects lead to the narrowness of solitonic
solutions [5]. Furthermore, it has been established that the
solitons are stable for $ p < 8 $ where $p$ corresponds to the power of
nonlinearity [6] (see, e.g., eqn. (1) below).
The analytic expression for the soliton
solution to the fifth order $KdV$($FOKdV$)-type equation
and two analytic solutions
to the fifth order $mKdV$$(FOmKdV)$-type equations have
also been found [7].\\

It is obvious that the above equations, their stability and solutions are of
considerable theoretical interests for the understanding of some physical
processes in the realm of fluid mechanics and plasma physics. \\

It is a well known fact that 1+1 dimensional (super)integrable
equations (e.g., (super)$KdV$ and (super)Boussinesq, etc.) play a
prominent role
in the understanding of (super)string theory, $W$-type (super)algebras,
statistical mechanics at critical temperature, $2D$ gravity theories,
fluid mechanics, etc. Their integrability properties stem from the
presence of an infinite number of commuting conserved quantities. The
existence of second Hamiltonian structure, zero-curvature representation,
Lax-pair formulation, multi-soliton solutions, Miura-type transformations,
etc.,
are also some of the key features of integrability (see, e.g., [8--10]).
The purpose of the present work is to study the highly nonlinear
$FOKdV$ and $FOmKdV$ equations in the light of the above integrability
properties as it is always worthwhile to look for a nonlinear
evolution equation that is integrable or partially integrable
[11]. To this end in mind,
we first develop a Lagrangian
formulation for the above dynamical equations following the procedure
adopted by Cooper etal.[12] in the context of ``compactons''
and derive conserved quantities
from the symmetries of the Lagrangian due to Noether theorem. We
discuss a consistent Hamiltonian formulation for the above systems and
concentrate on the Poisson bracket structures to study the commuting
properties of the conserved quantities.
It turns out that for a specific value of the power of
nonlinearity, the conserved quantities are recursively related to one-another.
However, the second Poisson bracket structure is found to be {\it not} a
consistent bracket for an integrable system [13]. The
$FOmKdV$-type equation is derived from the $FOKdV$-type equation by a
Miura-type transformation. It turns out that the spatial dependence
of the fields involved is such that the meaning of solitonic solutions
to these equations is lost. In contrast to the
usual $KdV$ and $mKdV$ equations which are related by Miura maps, this
result implies that $FOKdV$ and $FOmKdV$ equations are independent in their
own right. A consistent Lagrangian and Hamiltonian
formulations for the $FOmKdV$ equation is also developed.

The material of our work is organized as follows. In Sec. 2, we derive the
$FOKdV$
equation from the scalar field Lagrangian and discuss its
symmetry properties. Sec. 3 is devoted to the Hamiltonian structures
and discussion of the integrability properties.
We try to understand how far
and how much these properties can be discussed vis-a-vis the usual $KdV$
and $mKdV$ equations.  In Sec. 4, we derive $FOmKdV$ equation from the
starting $FOKdV$ equation by a Miura-type transformation and
demonstrate that
such type of relation entails the spatial dependence of the fields involved
to be some fractional power over spatial variable($x$). As a consequence,
it is
difficult to provide some physical interpretation to this result in the
language of solitonic solution.
Finally, in Sec. 5, we make
some concluding remarks and provide an outlook of the whole discussion.\\

\vspace{0.1cm}

\noindent
{\Large \bf 2 Lagrangian Formulation} \\

\vspace{0.1cm}

\noindent
We derive here the $FOKdV$-type equation
\begin{equation}
u_{t}
+ \;\gamma\; u_{5x}
+ \beta\; u_{3x} + \alpha \;u^p\; u_{x} = 0,
\label{1}
\end{equation}
from the first-order Lagrangian density $ {\cal L} $ that is
expressed in terms of
the derivatives ($\phi_{t}= \partial_{t} \phi, \; \phi_{x}
= \partial_{x} \phi$)
on a scalar field $\phi$. The continuous symmetries of the action
($ S = \int dt \int dx \;{\cal L} $)
lead to the derivation of three conserved
quantities due to Noether theorem.
We begin with the
following first-order Lagrangian
\begin{equation}
L(p) = \int dx {\cal L}
\equiv \int dx \Bigl [ \frac{1}{2} \phi_{x}\;\phi_{t}
+ \frac{\alpha\; (\phi_{x})^{ p + 2}}{ ( p + 1) ( p + 2) }
- \frac{\beta } {2} \phi_{2x}^2
+ \frac{\gamma} {2}\;\phi_{3x}^2 \Bigr ],
\label{2}
\end{equation}
where $ p > 0 $ and c-number parameters
$ \alpha, \beta, \gamma $ of (1) and (2) obey certain restrictions for
the physically meaningful solutions to the equation (1) in the context
of fluid mechanics and plasma physics (see, e.g., [4]).
The Euler-Lagrange equation
\begin{equation}
\frac{d} {dx} \bigl ( \frac{\delta L}{\delta \phi_{x}} \bigr )
+ \frac{d} {dt} \bigl ( \frac{\delta L}{\delta \phi_{t}} \bigr )
= 0,
\label{3}
\end{equation}
with
\begin{eqnarray}
\frac{\delta L}{\delta \phi_{t}}
&=&    \Pi_{\phi}  = \frac{\phi_{x}}{2} ,\nonumber\\
\frac{\delta L}{\delta \phi_{x}} &=& \frac{1}{2} \phi_{t}
+ \frac{\alpha\; (\phi_{x})^{ p + 1}}{ ( p + 1) }
+ \beta \; \phi_{3x}
+ \gamma \;\phi_{5x},
\label{4}
\end{eqnarray}
leads to the following dynamical equation of motion
\begin{equation}
 \phi_{xt} + \alpha\; (\phi_{x})^{ p }\; \phi_{2x}
+ \beta \; \phi_{4x}
+ \gamma \;\phi_{6x} = 0.
\label{5}
\end{equation}
The identification $ \phi_{x} = u (x,t) $ leads to the derivation of (1).

We would like to dwell a bit more on the symmetries of the above Lagrangian
as they provide a physical insight into the system described by
it. Since only derivatives are present in the above Lagrangian, it is
straightforward to check that the following transformation
with a global infinitesimal parameter $\varepsilon$
\begin{equation}
\delta_{1} \phi = 2 \varepsilon,
\label{6}
\end{equation}
is a symmetry transformation because $ \delta_{1} {\cal L} = 0 $. Furthermore,
the space and time translations $ x \rightarrow x - \varepsilon,
t \rightarrow t - \varepsilon $ lead to the following transformations on the
scalar field $\phi$ ( since $ \phi (x^\prime, t^\prime) = \phi (x,t)$ )
\begin{eqnarray}
\delta_{2} \phi &=&  \varepsilon\; \phi_{x},\nonumber\\
\delta_{3} \phi &=&  \varepsilon\; \phi_{t}.
\label{7}
\end{eqnarray}
These are also the symmetry transformations of the action $S$ as the above
Lagrangian undergoes the following change
\begin{eqnarray}
\delta_{2} {\cal L } &=&  \varepsilon\;{\cal L}_{x}
\equiv \frac{\partial}{\partial x} \bigl ( \varepsilon {\cal L} \bigr ),
\nonumber\\
\delta_{3} {\cal L} &=&
\varepsilon\; {\cal L}_{t}
\equiv \frac{\partial}{\partial t} \bigl ( \varepsilon {\cal L} \bigr ),
\label{8}
\end{eqnarray}
where $\varepsilon$ is a global infinitesimal parameter
(as was the case in (6)).\\

The continuous symmetry transformations (6) and (7) imply presence of
conserved quantities ($ I's $) due to well-known Noether theorem. These are
juxtaposed, with the identification  $ \phi_{x} = u $, as follows
\begin{eqnarray} I_{1} &=&  \int dx \phi_{x} \qquad \rightarrow  \qquad \int
dx u, \nonumber\\ I_{2} &=&  \frac{1}{2}\int dx \phi_{x}^2 \qquad \rightarrow
\qquad \frac{1}{2}\int dx u^2, \nonumber\\ I_{3} &=&  \int dx \Bigl (
\frac{\beta } {2} \phi_{2x}^2 - \frac{\alpha\; \phi_{x}^{ p + 2}}{ ( p + 1) (
p + 2) } - \frac{\gamma} {2}\;\phi_{3x}^2 \Bigr ),\nonumber\\ &\rightarrow&
 \int dx \Bigl ( \frac{\beta } {2} u_{x}^2
- \frac{\alpha\; u^{ p + 2}}{ ( p + 1) ( p + 2) }
- \frac{\gamma} {2}\;u_{2x}^2 \Bigr ).
\label{9}
\end{eqnarray}
Note that in the derivation of $I_{3}$, we have used
\begin{equation}
I_{3} = \int dx J_{0} \equiv \int dx
\Bigl ( \frac{\delta_{3} \phi}{\varepsilon}
\frac{\partial {\cal L}}{\partial \phi_{t}} - {\cal L} \Bigr ),
\label{10}
\end{equation}
as the Lagrangian transforms to the time derivative of itself under
global time translation (7). It can be easily seen that
the time translation generator ``energy'' ($I_{3}$) is nothing
but the Hamiltonian function ($ H (p)$) corresponding to the first-order
Lagrangian function (2). It becomes transparent  in the following Legendre
transformation
\begin{equation} H (p) = \int dx \bigl [ \Pi_{\phi} \; \phi_{t}
\bigr ] -  L(p) \equiv \; I_{3}.  \label{11} \end{equation}
The conservation laws for
the quantities $I_{1,2}$, corresponding to the ``area'' under $u$ and
``momentum'',
can be understood directly from the
form of the dynamical equation. It is obvious from (1) that $I_{1}$ is
a conserved quantity due to the fact that r.h.s. of this equation is a total
space derivative.  For the conservation law of the space translation
generator ($I_{2}$), the equation of motion (1), implies that
\begin{equation}
u u_{t} = - \frac{\partial}{\partial x} \Bigl [
\frac{\alpha\; u^{ p + 2}}{ ( p + 2) }
+ \beta \; ( u u_{2x} - \frac{1}{2} u_{x}^2 )
+ \gamma\; ( u u_{4x} - u_{x} u_{3x} + \frac{1}{2} u_{2x}^2 ) \Bigr ].
\label {12}
\end{equation}
In general, for an equation of the type (1) with any order of odd derivatives
on $u$ and a nonlinear term of the type $ u^p\; u_{x} $, it can be seen that
$ u\;u_{t}$ is always a total space derivative. It is because of the fact
that for $ n = 0, 1, 2, 3 ..............$
\begin{equation}
u_{(2n +1) x} \;u = \frac{\partial}{\partial x}
\Bigl [ \sum_{r=0, n > r}^{n-1} (-1)^{r}\; u_{(2n -r) x}\;u_{rx}
+ (-1)^n \frac{1}{2} (u_{nx})^2 \Bigr ]
\label{13}
\end{equation}
where $ u_{0 x} = u, r = 0,1,2,3,...............(n-1) $. This demonstrates
that the ``momentum'' $I_{2}$ is always a conserved quantity for the kind of
equation we are discussing.\\

From the above Hamiltonian function, one can derive the dynamical equation
(5) (and (1) with proper identification) by exploiting the following canonical
Poisson-bracket structure for the scalar field
\begin{equation} \{ \phi (x,t), \phi_{y}(y, t)
\}_{PB}^{(1)}\; = \delta (x-y), \label{14}
\end{equation}
in the Hamilton
equation of motion
\begin{equation} \phi_{t} = \{ \phi, H(p) \}_{PB}^{(1)}
\equiv -\gamma\;\phi_{5x} - \beta\;\phi_{3x}
- \alpha\;\frac{\phi_{x}^{p+1}}{p
+ 1}.  \label{15} \end{equation}
It is evident that the total space derivative
of the above equation is nothing but the dynamical equation (5) which, in
turn, takes the form of equation (1) with the identification
$ \phi_{x} = u $.
Similar operation on the Poisson bracket (14) leads to the Hamiltonian
structure on $ u$ (the first PB) as
\begin{equation} \{ u(x), u(y) \}_{PB}^{(1)}\; =
\frac{\partial}{\partial x} \delta (x-y). \label{16}
\end{equation}
Thus, we establish the fact that
the Lagrangian and Hamiltonian formulations for the scalar
field lead to the
derivation of the Poisson-bracket structure for the $u$ fields as has been
taken by Karpman in Ref. [4] for the Hamiltonian function $I_{3}$.
{\it It is
interesting to point out that all the conserved quantities of equation (9)
commute with one-another under the choice of the Poisson bracket (16)}.\\

\vspace{0.05cm}

\noindent
{\Large \bf 3 Second-Hamiltonian Structure and Recursion Relations } \\

\vspace{0.05cm}

\noindent
Here we demonstrate that for
$ p = 1$, these conserved quantities are recursively related to one-another
for any arbitrary values of $ \alpha, \beta, \gamma $ as
there exists a second
Poisson bracket for the equation of motion (1) with the Hamiltonian as the
``momentum'' expression $I_{2}$. Thus, to some extent,
the $FOKdV$-type
equation does mimic the usual $KdV$ equation.\\

We know that the $KdV$ equation is an integrable equation because it supports
an infinite number of conserved quantities that are recursively related to
one-another due to the existence of a ``consistent'' second Hamiltonian
structure [13]. The first- and the second brackets satisfy the famous Jacobi
identities.  Here, we try to see how much and how far, the $FOKdV$
equation can mimic the anologue of the recursion relations of $KdV$
equation. To this end, it can be seen that the following Hamilton equations
\begin{equation}
u_{t}
= \{ u, I_{3} \}^{(1)} =  \{ u, I_{2} \}^{(2)},
\label{17}
\end{equation}
with the first bracket (cf., equation (16))
\begin{equation}
 \{ u(x,t), u(y,t) \}^{(1)} = - D^{(1)} \;\delta (x-y),
\label{18}
\end{equation}
and the second bracket
\begin{equation}
 \{ u(x,t), u(y,t) \}^{(2)} = \;D^{(2)}   \; \delta (x-y),
\label{19}
\end{equation}
where,
\begin{eqnarray}
D^{(1)}&=& \partial \equiv \;\frac{\partial}{\partial y}, \nonumber\\
D^{(2)}&=& \bigl [ \gamma \;\partial^5 + \beta \;\partial^3 + C\;u^p \;\partial
+ \{ ( p + 1 )\; C - \alpha\; \} u^{p-1}\; u_{y} \bigr ] ,
\label{20}
\end{eqnarray}
lead to the
derivation of equation (1) for any arbitrary value of the constant $C$.
For the $FOKdV$-type equation, the analogue of $KdV$ type
recursion relations, with Poisson brackets (18) and (19), are
\begin{equation} \hat {D^{(1)}} \Bigl ( \frac{\delta
I_{k}}{\delta u} \Bigr ) = \hat {D^{(2)}}\; \Bigl ( \frac{\delta
I_{k-1}}{\delta u} \Bigr ) , \label{21} \end{equation}
where $ k \geq 2 $ and
opertators $ \hat {D^{(1,2)}} $ can be computed from the brackets (18) and
(19).  These operators for our discussion are
\begin{eqnarray} \hat {D^{(2)}} &=& - \bigl [
\gamma \;\partial^5 + \beta \;\partial^3 + C\;u^p \;\partial
- ( C  - \alpha ) u^{p-1}\; \partial u \bigr ] ,\nonumber\\
\hat {D^{(1)}} &=& \partial.
\label{22} \end{eqnarray}
For $ k = 2 $, the above recursion relation leads to
the following expression for $I_{2}$
\begin{equation} I_{2} = \frac{( C -
\alpha )}{p\; ( p + 1 )}\; \int dx \;u^{ p + 1}. \label{23} \end{equation}
However, the comparison with the starting momentum expression ( i.e.  $ I_{2}
= \frac{1}{2}\int dx \;u^2 $ of equation (9)), puts  certain restrictions on
$p$ and $C$ as given below:
\begin{equation} p = 1
\qquad \mbox{and}  \qquad    C = 1 + \alpha. \label{24}
\end{equation}
As a result, the operator $\hat {D^{(1)}}$ remains intact but the
operator $\hat {D^{(2)}}$ modifies as follows:
\begin{equation}
\hat {D^{(2)}}= - \bigl [ \gamma\; \partial^5 + \beta\; \partial^3
+  ( 1 + \alpha
)\;u \;\partial - \partial u \; \bigr ] .  \label{25} \end{equation}
It will
be noticed that the form of the equation (1) is changed because there is a
restriction on $p$ ( i.e. $ p = 1$ ) but c-number parameters
$\alpha$, $\beta$ and $\gamma$ remain intact as there are no restriction on
them due to the recursion relations.

We go a step further to compute $I_{3}$ from the above recursion relation
with the modified version of second Poisson bracket (19).
The corresponding recursion relation with the above
$ \hat {D^{(2)}} $ (cf. (25))
\begin{equation}
\frac{d}{dx} \Bigl ( \frac{\delta I_{3}}{\delta u} \Bigr )
= \hat {D^{(2)}}\; \Bigl ( \frac{\delta I_{2}}{\delta u} \Bigr ) ,
\label{26}
\end{equation}
leads to
\begin{equation}
I_{3} = \int dx \Bigl (  \frac{\beta } {2} u_{x}^2
- \frac{\alpha }{ 6 } \; u^3
- \frac{\gamma} {2}\;u_{2x}^2 \Bigr ).
\label{27}
\end{equation}
The comaprison with the ``energy'' expression $I_{3}$ (for $ p=1 $
in equation (9)), puts
no constraint on $\alpha, \beta, \gamma$ and all the three conserved
quantities are recursively related.
{\it It can be checked that, for $p=1$,
all of them commute
with one-another under the Poisson bracket structure (19) as well}.\\

The next order recursion relation should yield the next conserved quantity
$I_{4} $ if the second Hamiltonian structure is a ``consistent''
bracket of an
integrable equation.  In fact, the following expression comes out from
the next recursion relation for any arbitrary value of $ \alpha, \beta $
and $ \gamma $:
\begin{eqnarray}
\frac{\delta I_{4}}{\delta u}
&=& \gamma \bigl ( u_{2x}^2 + u u_{4x} - 2 u_{x} u_{3x} )
+ \frac{\alpha}{6} u^3 + \beta (u\;u_{2x} - u_{x}^2) + \beta^2 u_{4x}
+ \frac{\alpha^2}{3} u^3
\nonumber\\
&+& \gamma^2  u_{8x} + 2 \beta \gamma \;u_{6x}
+ \alpha \beta ( \frac{1}{2} u_{x}^2 + 2 u u_{2x} )
+ \alpha \gamma ( \frac{7}{2} u_{2x}^2 +2 u u_{4x} + 3 u_{x} u_{3x}).
\label{28}
\end{eqnarray}
To obtain the expression for $I_{4}$, one should see that the r.h.s.
is a total variation of some quantity.
It turns out that for the choice $ \alpha = 3 $, the r.h.s. reduces to
a simpler form as given below
\begin{eqnarray}
\delta I_{4}
&=& \Bigl [ 7 \gamma \bigl ( \frac{23}{14} u_{2x}^2
+ u u_{4x} + u_{x} u_{3x} \bigr )
+ 7 \beta (u\;u_{2x} + \frac{1}{2} u_{x}^2)\nonumber\\
&+& \beta^2
u_{4x} + \gamma^2  u_{8x} + \frac{7}{2}\; u^3 + 2 \beta
\gamma \;u_{6x} \Bigr ] \; \delta u. \label{29}
\end{eqnarray}
Except the first term, all the other terms are total variation on a certain
local quantity modulo some total space derivatives. This can be seen as 
follows:
\begin{eqnarray}
&& (\frac{1}{2} u_{x}^2 + u\;u_{2x}) \delta u = \frac{1}{4}
\delta  (u^2\;u_{2x})
\qquad
u_{8x} \delta u = \frac{1}{2} \delta (u_{4x}^2) \nonumber\\
&& u_{6x} \delta u = - \frac{1}{2} \delta (u_{3x}^2), \quad
u_{4x} \delta u =  \frac{1}{2} \delta (u_{2x}^2), \quad
 \frac{7}{2} u^3 \delta u = \frac{7}{8}\;\delta (u^4).
\label{30}
\end{eqnarray}
It is interesting to note that
the first term can not be cast as a total variation of a definite local
quantity. The argument runs as follows. There are only four candidates
with three $u$ fields and four derivatives on them
\begin{equation}
 u^2\;u_{4x}, u_{2x}^2\;u, u\;u_{x}\;u_{3x}, u_{x}^2\;u_{2x},
\label{31}
\end{equation}
whose variation and/or
the variation of their linear combination might produce the combination
of the first term. It turns out that the variation on the last term is zero
modulo some total space derivative term. All the other three candidates
produce the same result due to the variation on them. For instance, for one
of them the variation is
\begin{equation}
 ( 2 u\;u_{4x} + 4 u_{x}
u_{3x} + 3 u_{2x}^2 ) \delta u = \frac{1}{2}\;
\delta (u^2\;u_{4x}).  \label{32}
\end{equation}
Thus, the $FOKdV$ equation is {\it not} an integrable
equation because the second Hamiltonian structure is not consistent [13].
It can be
also seen that the second-Poisson bracket does not satisfy the Jacobi
identities (see, e.g., for more detail, Ref. [12] for a similar
kind of situation).\\

\vspace{0.05cm}

\noindent
{\Large \bf 4 Towards a Miura-type transformation}  \\

\vspace{0.05cm}

\noindent
In this Sec., we derive the $FOmKdV$-type equation (i.e., $ p = 2 $)
from the starting $FOKdV$ equation by Miura-type transformation
(see, e.g., [13]).
This equation is also of theoretical interests in the context of fluid
mechanics and plasma physics. We develop here the Lagrangian formulation
for the $FOmKdV$ and discuss in a nut-shell the Poisson bracket
structures for this system on similar lines as that of the $FOKdV$
equation ( cf. Sec. 2).  The solitonic solution to this equation has
been found in Ref. [7] where two analytic solutions have been obtained.

To start with, we know that for the usual $KdV$ equation, there exists
a transformation that relates the original $KdV$ equation to the $mKdV$.
The question we address is : Can one derive the fifth-order
$mKdV$ equation from the starting equation (1)? To answer this question
we make the following transformation
\begin{equation}
 u(x,t) = v_{x} (x,t) - \frac{\alpha}{6\beta} \; v^2 (x,t),
\label{33}
\end{equation}
in equation (1).
It will be noticed that the inclusion of higher order derivative terms
on the
r.h.s. of equation (33) leads to more complicated expressions which are
difficult to handle.
The ensuing equation is realized on $v(x,t)$ as
\begin{equation}
v_{t}  +
\;\gamma\; v_{5x} + \beta\; v_{3x}
- \frac{\alpha^2}{6 \beta} \;v^2\; v_{x} = 0,
\label{34} \end{equation}
which can be recognized as the  $FOmKdV$ and can be re-expressed in
the form of a conservation law as given below:
\begin{equation}
v_{t} +
\frac{\partial }{\partial x} \;\Bigl [ \gamma\; v_{4x} + \beta\; v_{2x} -
\frac{\alpha^2}{18\beta} \;v^{3} \Bigr ] = 0.  \label{35}
\end{equation}
It should be added, however, that for the validity of (35),
 the field $ v(x,t) $ has to satisfy an extra condition given by
\begin{equation}
v_{x}\; v_{4x} + 2\; v_{2x}\; v_{3x} = 0.
\label{36}
\end{equation}
We have to find a solution for the above restriction if Miura transformation
(33) has to make some sense. The equation (36) can be re-written as a total
space derivative of a certain quantity, given by:
\begin{equation}
\frac{d}{dx} \bigl [ v_{x}\; v_{3x} + \frac{1}{2} \; ( v_{2x} )^2 \bigr ]
= 0. \label{37}
\end{equation}
The quantity in the square bracket has to be a constant. Choosing the
constant to be zero, we obtain the $x$ dependence of $ v $ as
\begin{equation}
v(x) = A\; x^{5/3},
\label{38}
\end{equation}
for an arbitrary constant $ A $. It will be
noticed that equation (38) is a nontrivial solution for the Miura
transformation and there is no constraint on the $t$-dependence. It is also
evident that for such a choice ($ v(x) \sim x^{5/3}$), the spatial dependence
of $u$ field will also change accordingly. Thus, there is a problem about
solitonic solutions for such kind of restriction.
It is well known that for the usual $KdV$ equation,
no such restrictions as (36--38)
arise. For instance, if $\gamma = 0 $ and
$ p = 1 $ in the starting equation (1), there
will be no restrictions like (36--38) and for any arbitrary
$\alpha, \beta$, we have
\begin{eqnarray}
&& u_{t} + \beta\;u_{3x} + \alpha\;u\;u_{x} = 0, \nonumber\\
&& u(x,t) = v_{x} (x,t) - \frac{\alpha}{6\;\beta} \; v^2, \nonumber\\
&& v_{t} + \beta\;v_{3x} -\frac{\alpha^2}{6 \beta} \;v^2\;v_{x} = 0,
\label{39}
\end{eqnarray}
as $KdV$ equation, Miura transformation and the $mKdV$ equation.\\

We conclude from equation (38) that $FOKdV$ and $FOmKdV$ equations are
independent in their own right. There is no Miura type of mapping
which can relate these equations preserving the nature of stationary
solitonic solutions that have been obtained in  Ref. [7].\\

The $FOmKdV$ equation of motion (34) can
be derived  from the
following first-order Lagrangian
\begin{equation}
L = \int dx \Bigl [ \frac{1}{2} \phi_{x}\;\phi_{t}
- \frac{ \alpha^2 } { 72\;\beta} (\phi_{x})^4
- \frac{\beta } {2} (\phi_{2x})^2
+ \frac{\gamma} {2}\;(\phi_{3x})^2 \Bigr ).
\label{40}
\end{equation}
The Euler-Lagrange equation, emerging from the invariance of the action
($ \delta S = 0 $), leads to the following dynamical equation:
\begin{equation}
 \phi_{xt} - \frac{\alpha^2}{6\;\beta} (\phi_{x})^{ 2 }\; \phi_{2x}
+ \beta \; \phi_{4x}
+ \gamma \;\phi_{6x} = 0.
\label{41}
\end{equation}
The identification $ \phi_{x} = v (x,t) $ leads to the rederivation of
(34). The anologue of the symmetry transformations (6) and (7) for the
above Lagrangian lead to the following conserved currents due to the
Noether theorem
\begin{eqnarray}
J_{1} &=& \int dx \;v,  \qquad         J_{2} = \frac{1}{2}\int dx \;v^2,
\nonumber\\
J_{3} &=& \int dx \Bigl (
 \frac{ \alpha^2}{ 72\;\beta }\; v^4
+\frac{\beta } {2} v_{x}^2
- \frac{\gamma} {2}\;v_{2x}^2 \Bigr ).
\label{42}
\end{eqnarray}
The equation of motion (34) can be derived from the Hamilton equation
as well. The corresponding Hamiltonian functions  are none other than
$ J_{3} $ and $ J_{2} $ as illustrated below:
\begin{equation}
v_{t} = \{ v, J_{3} \}^{(1)} =  \{ v, J_{2} \}^{(2)},
\label{43}
\end{equation}
where the first bracket  and the second brackets are
\begin{eqnarray}
\{ v(x,t), v(y,t) \}^{(1)} &=& - \frac{\partial}{\partial y} \delta (x - y),
\nonumber\\
\{ v(x,t), v(y,t) \}^{(2)} &=&
 \bigl [ \gamma \partial^5 + \beta \partial^3 + C v^2 \partial
+ ( 3\;C + \frac{\alpha^2}{6\;\beta} ) v_{y}\;v \bigr ] \delta (x-y),
\label{44}
\end{eqnarray}
for arbitrary constant  $C$ and
$ \partial = \frac{\partial}{\partial y} $.  \\

\vspace{0.05cm}

\noindent
{\Large \bf 5 Discussions} \\

\vspace{0.05cm}

\noindent
The integrability properties play a pivotal role in furnishing the wealth
of informations on the  rich mathematical
structure encoded in a given nonlinear evolution equation. This, in turn,
leads to gain some deep physical insight into the  phenomena that are
described by these equations. In the light of this argument, we have studied
the $FOKdV$ and $FOmKdV$ equations.
We have developed a Lagrangian formulation for the $FOKdV$ equation
and obtained
the conserved quantities from the symmetry properties of this Lagrangian. We
have shown that these conserved quantities are recursively related to
one-another for $ p = 1$ due to the existence of bi-Hamiltonian structures.
The recursion relations do not lead to the determination of more than three
conserved quantities. Thus, as it appears, the $FOKdV$ equation is not a
completely integrable system.\\

The $FOKdV$ equation mimics some of the key features of the partially
integrable system. For instance, the soliton solutions are stable and
preserve their shape for considerably large values of the power of
nonlinearity. There exists a Miura type transformation that relates $FOKdV$
and $FOmKdV$ equations for the specific space-dependence of the fields on
which these equations are realized. However, from the point of view of the
solitonic solutions, this dependence is not an interesting feature. This
leads to the conclusion that
$FOKdV$ and $FOmKdV$ equations are independent in their own right and there
exists no Miura-type mapping which can relate these equations without
spoiling the solitonic solutions to these equations.
Furthermore, for $ p = 1 $, the conserved quantities are
recursively related to one-another (due to the existence of two Poisson
bracket structures). One can also see that the $FOKdV$ and $FOmKdV$ equations
can be expressed as an Abelian-type zero-curvature representation
($F_{tx}=0$)\\
\begin{equation}
F_{tx} = \partial_{t} A_{x} - \partial_{x} A_{t}
\equiv \bigl [ \frac{\partial}{\partial t } + A_{t},
\frac{\partial}{\partial x} + A_{x} \bigr ] = 0,
\label{45}
\end{equation}
where gauge connection $A_{x}$ and $A_{t}$ can be read off from the
$FOKdV$ and $FOmKdV$ equations respectively, as listed below\\
\begin{eqnarray}
A_{x} &=& u (x,t) \qquad \mbox{and} \qquad v(x,t),\nonumber\\
A_{t} &=& - ( \gamma u_{4x} + \beta u_{2x} + \alpha
\frac{u^{p +1}} { p + 1})\quad  \mbox{and} \quad
\frac{\alpha^2}{18 \beta} v^3 - \beta v_{2x} - \gamma v_{4x}.
\label{46}
\end{eqnarray}

It will be an interesting venture to study the integrability properties of
these equations where we can use the Painleve test [11] and consider the
multi-soliton solution by exploiting the Hirota's bilinear form and its
generalization [14]. It is worthwhile to mention that in Ref. [14],
it has been claimed that the existence
of three solitonic solutions imply integrability. These are the issues
for future investigations and further works have to be done
in this direction.\\

\vspace{0.2cm}

\noindent
{\Large \bf Acknowledgements}  \\

\vspace{0.05cm}

\noindent
Many constructive criticisms and comments
by A. Khare are gratefully acknowledged. It is a pleasure
to thank J. Hietarinta, V. Mel'nikov and P. Zhidkov
for some fruitful and stimulating discussions.\\

\end{document}